\documentstyle[NATO,epsf,epsfig,namedreferences]{CRCKAPB} 

\newcommand{\Msolar}{\mbox{\,$\sf M_{\odot}$}}        

\begin{opening}
\title{Subluminous B Stars and Progenitors of Helium Core White Dwarfs}
\author{U. HEBER}
\institute{Dr. Remeis-Sternwarte, Universit\"at Erlangen-N\"urnberg,\\ 
	   Sternwartstr. 7, D--96049 Bamberg, Germany}
\end{opening}
\begin{document}
\section{Introduction}

It is now generally accepted that the subluminous B (sdB) 
stars can be identified with 
models for Extreme Horizontal Branch (EHB) stars burning He in their core, 
but with a very tiny inert hydrogen envelope (Heber, 1986).
An EHB star bears great resemblance to a helium main-sequence star 
of half a solar mass and its further evolution should proceed similarly 
(i.e. directly to the white dwarf graveyard) 
as confirmed by evolutionary calculations (Dorman et al. 1993).
How stars evolve to the EHB configuration is controversial. 
The problem is how the mass loss mechanism in the
progenitor manages
to remove all but a tiny fraction of the hydrogen envelope at {\em
precisely} the same time as the He core has attained the minimum
mass ($\approx0.5$M$_\odot$) required for the He flash. 

There is growing evidence that close binary evolution is an important if 
not the dominant formation path of sdB stars (Maxted et al. 2001a, Heber et 
al. 2002). 
SdB stars can result from stable Roche lobe overflow or 
common envelope ejection models (Han et al. 2002). The companions are 
either low mass main sequence stars or white dwarfs.
SdB stars may also result from the merger of two He white dwarfs
(Webbink, 1984).

\section{Masses of sdB stars}

According to evolution theory the mass of an sdB star is fixed by the
core mass of the giant progenitor at the core helium flash 
($\approx$0.46$\ldots$0.48\Msolar). The envelope ($<$ 0.02\Msolar) does not 
contribute to the total mass significantly. Therefore half a solar mass is 
generally assumed to be the appropriate mass of an sdB star. 
A binary system containing an sdB star may offer the opportunity to measure
the mass of sdB star using Keplers third law from light and radial 
velocity curves. A direct solution for the masses of both components is 
possible for systems which are eclipsing and for which both radial 
velocity curves can be measured.
Although many subluminous B stars in binaries are known, 
no such system has yet 
been found. 

Today orbital periods have been determined for 38 sdB stars 
(Morales-Rueda et al. 2002 and 
references cited therein). In all cases the companions are invisible. 
For thirteen systems the companions have been identified as white dwarfs and 
only in five systems as low mass main sequence stars. 
Atmospheric parameters are available for 36 of them and are consistent with 
the predictions of evolutionary EHB models (Fig. 1a). 
Eclipses have been reported only for three of them: HW~Vir, PG~1336-018 and
HS~0705+6700. In all three cases the companions are low mass main sequence 
stars.
Hence the mass function and the inclination angle can be 
determined. However, the masses of both components can be derived if 
the gravity of the sdB stars can be determined accurately from spectroscopy.  
The analysis of the light and radial velocity curve combined with 
the spectroscopic gravity measurement results in sdB masses of 
0.54\Msolar (HW~Vir), 
0.5\Msolar (PG~1336-018) and 0.48\Msolar (HS~0705+6700) 
close to the canonical 
value of half a solar mass (see Drechsel et al. 2001). 

Even in the case of a non-eclipsing system we may be able to determine the mass.
If the system is sufficiently close, the rotation of the sdB star will be
tidally locked to the orbit. The projected rotational velocity can be 
measured from high quality spectral line profiles and therefore the 
inclination angle can be derived. For HS~0705+6700 the 
inclination derived form v\,sin\,i\ is consistent with that derived from the 
light curve analysis. The method has also been applied successfully to 
the non-eclipsing system PG~1017$-$086 (Maxted et al. 2001b).

\section{HD~188112}

HD~188112 is a bright (V=10.2) nearby sdB star. Its distance is well determined by 
a parallax measurement of the HIPPARCOS satellite to be 
d=81$^{+13}_{-11}$pc. The spectral analysis is based on 
high resolution (0.1\AA) optical spectra taken at the ESO 1.5m equipped 
with the FEROS spectrograph. Atmospheric parameteres are derived by 
matching the observed by synthetic spectra calculated from 
LTE model atmospheres. The results are T$_{\rm eff}$=19300$\pm$500\,K,
log~g = 5.51$\pm$0.1 and a helium content of only He/H = 5$\times$10$^{-5}$. 
These parameters are unusual for an sdB and place the star {it below}
the zero age Horizontal Branch (see Fig.~1a). It cannot have evolved off 
the EHB either, since the evolution of the EHB stars 
leads to increasing temperatures and lower gravities than observed for 
HD~188112 (see Fig.~1a).

\begin{figure}
\begin{center}
\vspace*{6cm}
\includegraphics{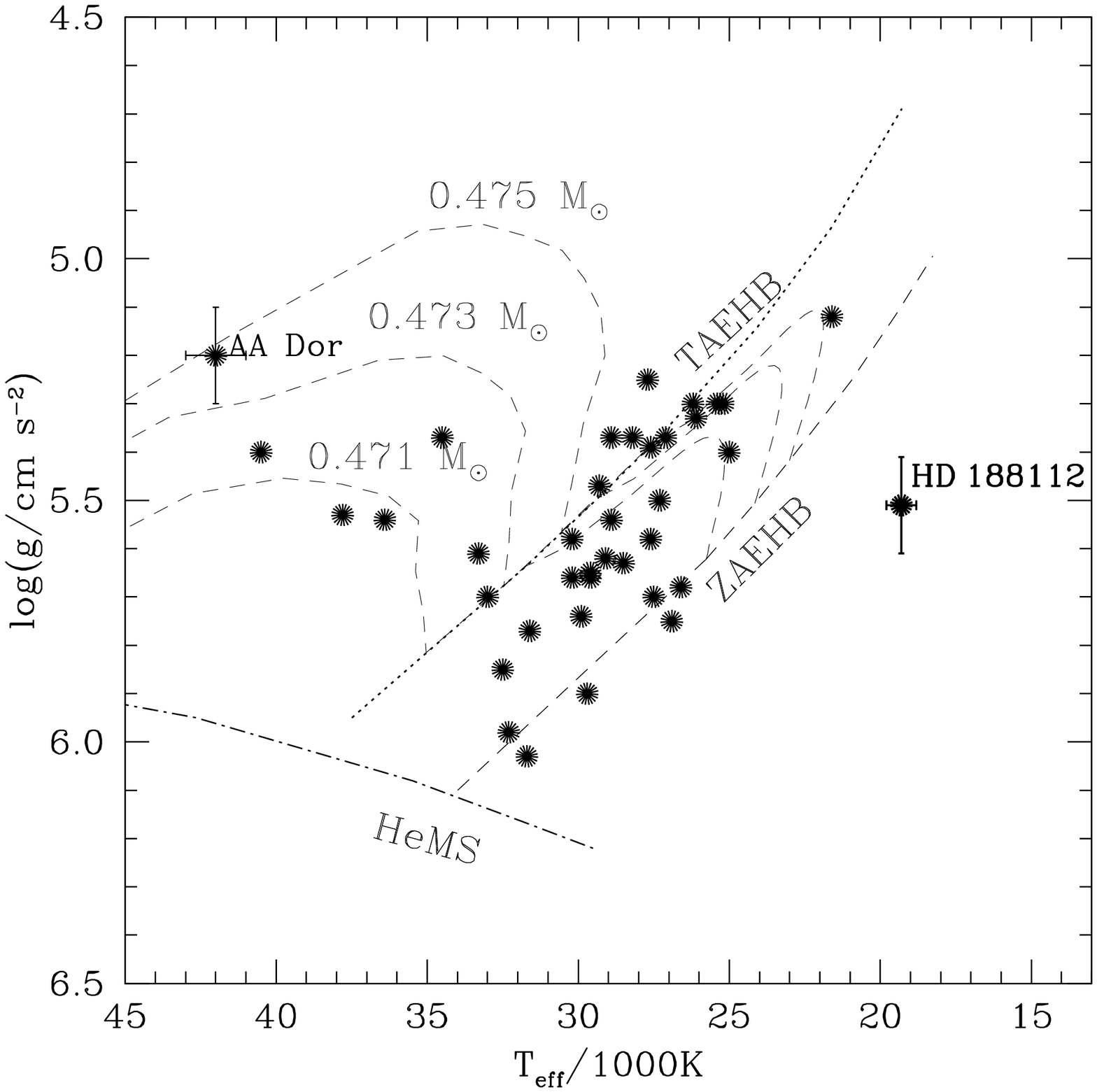}
\includegraphics{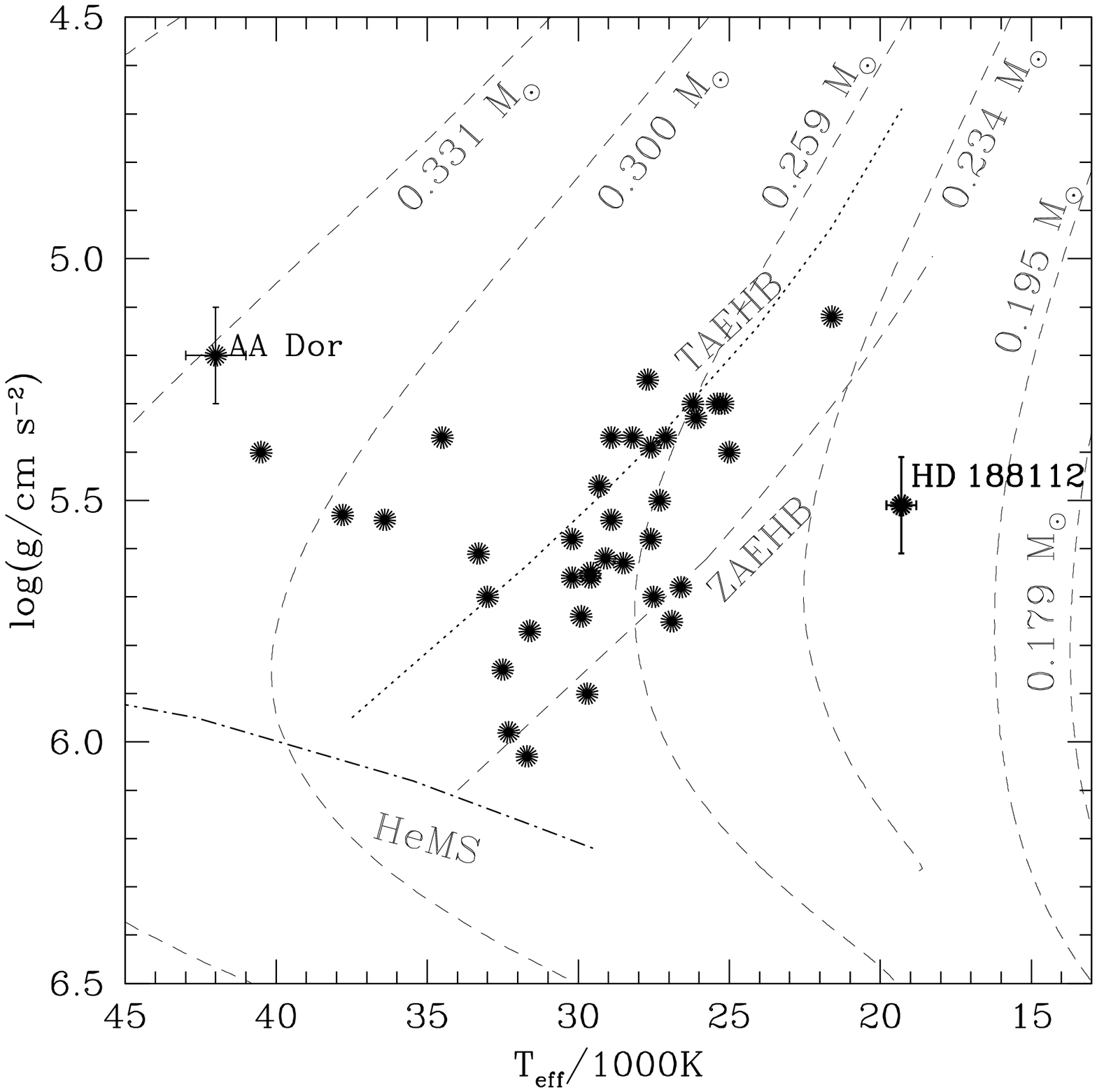}
\caption{a.) Left hand side: comparison of HD~188112 and AA~Dor in the 
(T$_{\rm eff}$, log~g)-plane to 
tracks for EHB evolution (Dorman, 1993). The tracks are labeled with the 
stellar mass.
b.) Right hand side: comparison of HD~188112 and AA~Dor 
(T$_{\rm eff}$, log~g)-plane to post-RGB tracks (Driebe et al., 1998). 
The tracks are labeled with the
stellar mass.}
\end{center}
\end{figure}

Combining the atmospheric parameters with the parallax measurement 
the mass of HD~188112 can be derived to be 
M~=~0.23$^{+0.15}_{-0.09}$\Msolar, far below the canonical sdB mass.
Therefore HD~188112 cannot be a core helium burning star. 
Such a low mass
object, however, can be formed in a close binary 
system when the progenitor star fills its Roche lobe on the first giant 
branch (RGB) well before the core mass has increased to the 
critical mass for the 
core helium flash. The star will then evolve to become a 
helium core white dwarf. The evolution of such post-RGB stars has been 
calculated by 
Driebe et al. (1998) and evolutionary tracks for different masses are shown
in Fig.~1. The position of HD~188112 in the (Teff, log~g)-diagram is 
bracketed by the evolutionary 
tracks for M=0.195\Msolar\ and M=0.234\Msolar;
 interpolation gives 0.22\Msolar,
in perfect agreement with the parallax based mass estimate.
The evolutionary time since departure from the RGB is of the order 
10$^8$ years according to the models of Driebe et al. (1998).   

If HD~188112 resulted from close binary evolution, the companion should
be detectable from the spectral energy distribution or from radial velocity 
variations. The optical spectrum does not show any hint for a second 
spectrum. Combining optical photometry with UV fluxes measured by the 
IUE satellite and NIR fluxes from the 2MASS catalog we can 
construct the spectral energy distribution. The latter is well 
matched by the sdB model flux distribution. 
Hence there is no evidence for light from the companion.  
However, HD~188112 was indeed found to be radial 
velocity variable. In August 2002 we measured the radial velocity 
curve 
with the FEROS spectrograph at the ESO 1.5m telescope. The preliminary 
analysis reveals a period of 0.6066 days and an half amplitude of K=188.4km/s
(Heber et al., in prep.). 
From the 
mass function we derive a lower limit to the companion mass of 
0.72\Msolar.  
Hence the companion cannot be a main sequence star but must be a 
compact object, most likely a white dwarf, but we cannot rule out 
a neutron star companion. The mass of the sdB is rather low for 
a helium core white dwarf. Such low mass white dwarfs, however, have 
been observed as companions to neutron stars.
The projected rotational velocity will be determined to constrain the
inclination angle and, therefore, improve the estimate for the lower 
limit to the companion mass. A measurement of its light curve would also be 
very valuable.


\section{Progenitors of helium core white dwarfs}

HD~188112 was found to be a bona-fide progenitor of a helium core white 
dwarf. Only few other candidates are known, e.g. HZ~22 (Sch\"onberner 1978), 
EGB~5 (Karl et al. these proceedings) 
and AA~Dor (Rauch, 2000). Unlike HD~188112 these stars lie 
above the EHB in the (T$_{\rm eff}$, log~g)-plane
as can be seen from Fig. 1a for AA~Dor. Hence their position can also be 
matched by post-EHB tracks.
Alternatively, the position of AA~Dor can be compared to post-RGB 
evolutionary tracks (Fig. 1b). Hence in the absence of a mass determination 
its evolutionary status cannot be assigned uniquely.

AA~Dor is an eclipsing single lined binary consisting of a hot sdOB star
(T$_{\rm eff}$=42000K, log~g=5.2) and 
a low mass main sequence star or brown dwarf (Rauch, 2000). From the comparison to 
post-RGB tracks (Fig. 1b) one can read of the mass as 0.33\Msolar. However, this
result is not consistent with the mass function and light curve (see Fig. 
17 of Rauch, 2000).

HD~188112 remains the most compelling case of a progenitor for a helium 
core white dwarf.


\end{document}